\documentclass[12pt,a4paper,twoside]{article}
\newcommand{\eq}[2]{\begin{align}\label{#1}#2\end{align}}

\newcommand{\nn}{\nonumber}
\newcommand{\pa}{\partial}
\newcommand{\ben}{\begin{enumerate}}\newcommand{\een}{\end{enumerate}}
\newcommand{\Ref}[1]{(\ref{#1})}

\newcommand{\td}{thermodynamic~}
\newcommand{\elm}{electromagnetic~}

\newcommand{\ga}{\gamma}
\newcommand{\ep}{\varepsilon}
\newcommand{\om}{\omega}\newcommand{\Om}{\Omega}

\usepackage{graphicx}
\usepackage{amsmath}\usepackage{bbm,bm,amsfonts,amssymb}

\begin{document}
\title{Free energy and entropy for thin sheets}
\author{M. Bordag\thanks{bordag@uni-leipzig.de}\\
\small Universit{\"a}t Leipzig}
\date{\today}

\maketitle


\begin{abstract}
We calculate the entropy and the temperature dependent part of the free energy for a free standing plane plasma sheet and for the a free standing plane slab of finite thickness with dispersion described by the plasma model. In case the plasma sheet describes a charged fluid, the entropy is positive. In case it describes a polarizable dipole sheet, the entropy takes negative values in a certain parameter region. Negative entropy is also observed for the slab, however with the reservation that the contribution from the surface plasmons was not accounted for.
\end{abstract}\thispagestyle{empty}

\section{\label{T1}Introduction}
There is a recent interest in calculation of free energy and entropy in Casimir-effect like configurations. As for the free energy, it is its contribution to the stability of thin films, which was recently investigated in \cite{klim17-95-012130} and \cite{klim17-29-275701}. However, there only the thickness dependent part was calculated, for which one can use the well-known Lifshitz formula. As for the entropy, recently  the question on the sign of the entropy for a free standing flat plasma sheet \cite{para17-96-085010} and sphere \cite{milt17-96-085007} were raised. The point is that negative entropies were repeatedly reported for Casimir-like situation, first in \cite{geye05-72-022111}, later in \cite{brev06-8-236} and \cite{khus12-45-265301},   where however only the thickness dependent parts were considered.
The complete entropy for a plasma sphere was recently calculated in \cite{bord1805.11241} and a region where it takes negative values was found.

In view of these developments we see an interest  to calculate the free energy and the entropy for   single sheets.  The first candidate is a plasma sheet which was investigated quite in detail in \cite{BV} and, as concerns its spectral properties, in \cite{bord05-38-11027}. This model has a relation to  2d electron gas \cite{fett73-81-367}. It is very simple and allows for quite explicit formulas. The next candidate is a plane dielectric slab of finite thickness. It is one of the basic models for the Casimir effect  (for details see \cite{BKMM}). For this configuration  the thickness dependent part of the free energy was recently investigated in \cite{klim17-95-012130} and \cite{klim17-29-275701} using the Lifshitz formula.

In general, there are two ways to calculate the free energy. One is the use of the Matsubara representation and the other uses the representation in terms of real frequencies involving the Boltzmann factor. The relation between them is described, for example, in Sect. 12.1.2 in \cite{BKMM}.
The zero temperature part of the free energy, i.e., the vacuum energy, is typically accompanied by ultraviolet divergencies. Their treatment is by now well known, but still somehow annoying. A fundamental advantage of the Lifshitz formula as describing only the thickness dependent part of the vacuum and free energies is that it is free of ultraviolet divergences. The same holds true for the temperature dependent part of the free energy and for the entropy, as defined by the temperature derivative from the free energy. Therefore, especially the calculation of the entropy does not need for any regularization and no related ambiguities should occur. We mention that the Matsubara representation of the free energy includes the vacuum energy and therefore also the ultraviolet divergencies. In this case one has to start with a regularization which one has to get rid of at the end. A   convenient way to do that is to use the Abel-Plana formula for transforming the sum over the Matsubara frequencies into an integration over real frequencies. In doing so, the vacuum energy separates and one is left with convergent expressions.  In the present paper we use this method. We calculate the temperature dependent part of the free energy and the entropy for the two mentioned models.

Throughout the paper we use units with $k_{\rm B}=\hbar=c=1$. We use the notation 'TX' in case a formula is valid for   both, 'TE' and 'TM' polarizations.

\section{\label{T2}Basic formulas}
In this section we introduce our basic formulas and the models considered. Almost all formulas can be found in literature, but we collect them to make the present paper largely selfcontained.
\subsection{\label{T2.1}Free energy and entropy}
We start with eq.(5.15) in \cite{BKMM} for the free energy of a system with eigenfrequencies $\om_J$,
\eq{1.1}{{\cal F}&=\frac{T}{2}\sum_{l=-\infty}^{\infty} \sum_J
    \ln\left(\xi_l^2+\om_J^2\right),
}
which represents the free energy at temperature $T$ as it can be derived, for example,  from the corresponding functional integral and where $\xi_l=2\pi T l$ are the Matsubara frequencies. This expression has  ultraviolet divergences and can be regularized in the following way,
\eq{1.2}{{\cal F}&=-\frac{\pa}{\pa s}
\frac{T\mu^{2s}}{2}\sum_{l=-\infty}^{\infty} \sum_J
    {\left(\xi_l^2+\om_J^2\right)^{-s}}_{|_{s=0}},
}
where $\mu$ is an arbitrary parameter having dimension of mass which can be introduced along with the regularization.
Next we transform the sum over the Matsubara frequencies into an integration using the Abel-Plana formula and get with
\eq{1.3}{{\cal F}&=E_0(s)+\Delta_T{\cal F}
}
the free energy split into its zero temperature part (vacuum energy),
\eq{1.4}{E_0(s) &= -\frac{\pa}{\pa s}
\frac{ \mu^{2s}}{2}\int_0^\infty\frac{d\xi}{\pi} \sum_J
    {\left(\xi^2+\om_J^2\right)^{-s}}_{|_{s=0}}
}
and the temperature dependent part
\eq{1.5}{\Delta_T{\cal F} &= -\frac{\pa}{\pa s}\sum_J
\frac{1}{\pi}\int_\ga\frac{d\om}{e^{\om/T}-1}
i\left(   \left(\om_J^2-(\om-i0)^2\right)^{-s}
        -\left(\om_J^2-(\om+i0)^2\right)^{-s} \right),
}
where the patch $\ga$ encircles the positive real half axis. This path can be tightened to the axis,
\eq{1.6}{\Delta_T{\cal F} &= -\frac{\pa}{\pa s}\sum_J
\frac{1}{\pi}\int_{\om_J}^\infty\frac{d\om}{e^{\om/T}-1}
 {2\sin(\pi s)} \left(-\om_J^2+\om^2\right)^{-s}.
}
Since this integral is obviously converging we may carry out the derivative and put $s=0$,
\eq{1.7}{\Delta_T{\cal F} &= \sum_J
\int_{\om_J}^\infty\frac{d\om}{e^{\om/T}-1},
}
and finally carry out the integration,
\eq{1.8}{\Delta_T{\cal F} &= \sum_J T\ln\left(1-e^{-\om_J/T}\right).
}
We arrived, for a fixed $J$, at the well known formula for the free energy of a single bosonic harmonic oscillator.

We proceed by assuming that we have two translational invariant directions parallel to the sheet with two dimensional momentum $\bm{k}=(k_1,k_2)$  ($k=|\bm{k}|$) and a one dimensional scattering setup in the direction perpendicular to the plane. In that case
the eigenfrequencies are
\eq{1.9}{\om_J=\sqrt{k^2+p_J^2},
}
where the $p_J$ result from a one dimensional scattering setup, for example like the one considered in \cite{bord95-28-755}. Dropping the empty space contribution, the sum in \Ref{1.8} can be transformed into
\eq{1.10}{\Delta_T{\cal F} &=\int\frac{d\bm{k}}{(2\pi)}
\left[
 T\ln\left(1-e^{-\om_{sf}(k)/T}\right)
+ \int_0^\infty\frac{dp}{\pi} T\ln\left(1-e^{-\om/T}\right)\frac{\pa}{\pa p}\delta(p)
\right],
}
with $\om=\sqrt{k^2+p^2}$ and where $\om_{sf}(k)$ is the frequency of the surface mode which is present in the TM-polarization. Further in \Ref{1.10}, $\delta(p)$ is the scattering phase shift. It can be expressed by
\eq{1.11}{\delta(p)&=\frac{1}{2i}\ln\frac{t(p)}{t(p)^*}
}
in terms of the transmission coefficient $t(p)$ and its complex conjugate.
We mention that $\frac{\pa}{\pa p}\delta(p)$ has also the meaning of the density of states.

The entropy $S$ can be obtained from the free energy by the \td formula
\eq{1.11a}{S&=- \frac{\pa}{\pa T}{\cal F}=\int\frac{d\bm{k}}{(2\pi)}
    \left[g\left(\frac{\om_{sf}(k)}{T}\right)
    +\int_0^\infty\frac{dp}{\pi} g\left(\frac{\om}{T}\right) \frac{\pa}{\pa p}\delta(p)\right],
}
where we introduced the notation
\eq{1.11b}{g(x)&=\frac{x}{e^x{-1}}-\ln\left(1-e^{-x}\right)
}
which in \Ref{1.11a} carries the temperature dependence. We mention that this function and the logarithm in \Ref{1.10} are exponentially decreasing for large argument which makes the temperature dependent part of the free energy and the entropy free of ultraviolet divergences.

Expressions \Ref{1.10} for the free energy and \Ref{1.11a} for the entropy are yet not the final ones. The reason is in their behavior at high temperature. As well known \cite{dowk78-11-895}, see also \cite{BKMM}, eq. (5.51),   for $T\to\infty$ the expansion
\eq{1.18}{{\cal F}&=
        -\frac{\zeta(3)a_{\frac12}}{4\pi^{3/2}}
                        \frac{(k_{\rm B}T)^3}{(\hbar c)^2}
        -\frac{a_1}{24} \frac{(k_{\rm B}T)^2}{\hbar c }
        -\frac{a_{\frac32}}{(4\pi)^{3/2}}k_{\rm B}T\ln(k_{\rm B}T)
        +O(T\ln T).
}
holds, where the $a_k$ are the heat kernel coefficients. These are well known, especially in the calculation of vacuum energy (for reviews see \cite{vass03-388-279} and \cite{kirs01b}) and describe the ultraviolet divergencies of the vacuum energy. For the model considered in this paper the coefficients entering \Ref{1.18} are non zero.  For example, for the plasma sheet these were calculated in \cite{bord05-38-11027} (eqs. (4.14) and (4.28)).
The contribution with the  coefficient $a_0$ results from the empty space (it describes the black body radiation) and is not present in our formulas after we dropped the empty space contribution in formula \Ref{1.10}.

In \Ref{1.18} we restored  for a moment the dependence on $k_{\rm B}$, $\hbar$ and $c$ (usually $\hbar$ is included in the coefficients) in order to demonstrate that the contributions displayed there are unphysical because of the inverse powers of $\hbar$ (see also the discussion in Sect. 5.1 in \cite{BKMM}). These contributions must be subtracted. Such subtraction may be considered as part of the ultraviolet renormalization which is needed for the zero temperature part of the free energy, i.e., for the vacuum energy, anyway. As known, the contributions form the coefficients $a_k$ with $k\le 2$ must be subtracted. The renormalization procedure allows for a freedom of a finite renormalization with the same set of heat kernel coefficients. Frequently, this freedom is fixed by a normalization condition.  We use this freedom to subtract the contributions growing in $T$ which are shown in \Ref{1.18} and have $\hbar$ in their denominators. This way we make sure that after subtraction the free energy tends for $T\to\infty$ to the classical limit which is linear in $T$, including possibly terms proportional to $T\ln(T)$. This way, the classical limit may be considered as a kind of normalization condition. Thus we define
\eq{1.19}{{\cal F}^{\rm high}&=
        \frac{(k_{\rm B}T)^3}{(\hbar c)^2}a_\frac12
        +\frac{(k_{\rm B}T)^2}{\hbar c }a_1,
\ \
    S^{\rm high}=
       -3k_{\rm B}\frac{(k_{\rm B}T)^2}{(\hbar c)^2}a_\frac12
        -2k_{\rm B}\frac{k_{\rm B}T}{\hbar c }a_1,
}
as high temperature parts and consider the differences
\eq{1.20}{{\cal F}^{\rm subtr} &= {\cal F}-{\cal F}^{\rm high}, \ \
            S^{\rm subtr} = S-S^{\rm high},
}
as the physical free energy and entropy of the considered systems.

\subsection{\label{T2.2}Models for flat sheets}
%
\subsubsection{\label{T2.2.1}Thin plasma sheet}
First we consider an infinitely thin sheet with in-plane polarizability. It can be thought as a two dimensional distribution of oscillators allowed to vibrate in the plane. In \cite{bart13-15-063028} it was called a {\it monoatomically thin insulator polarizable perpendicularly}. It is characterized by a frequency dependent plasma frequency
\eq{3.1}{\Om&= \Om_0\frac{ \om^2}{\om^2-\om_0^2+i0}
}
where
\eq{3.2}{\Om_0=\frac{4\pi e^2\rho}{m}
}
is a plasma frequency. Here $e$ is the charge and $m$ is the mass of the oscillators which are present with a two dimensional density $\rho$. We call $\Om_0$ plasma frequency since for $\om_0=0$ we get a model for a charged fluid
in the sheet. For more details we refer to \cite{BV}, where $\Om_0$ is called $q$.  We mention that the fluid needs for a immovable homogenous neutralizing background to avoid Coulomb interaction.

The dipoles as well as the fluid have a nonrelativistic dynamics and the usual coupling to the electromagnetic field. The resulting Maxwell equations remain unchanged outside the sheet, but are supplemented by matching conditions across the sheet. The details can be found in \cite{BV} and \cite{bart13-15-063028}. For $\om_0=0$ this model is also called {\it hydrodynamic model} and mimics to some extend the interaction of the \elm field with the $\pi$-electrons in graphene, whereby it must be mentioned that  the Dirac model gives a much more accurate description as it was shown in \cite{klim15-91-045412}.

The scattering setup in direction perpendicular to the sheet results in transmission coefficients
\eq{1.21}{t_{\rm TE}&=(1+Q_{\rm TE})^{-1}\ \  \mbox{with} \ \ Q_{\rm TE}=\frac{i\Om}{p},
\nn\\   t_{\rm TM}&=(1+Q_{\rm TM})^{-1} \ \ \mbox{with}\ \  Q_{\rm TM}=\frac{i\Om p}{\om^2},
}
for the two polarizations of the \elm field. The momenta are $\bm{k}$ in direction parallel to the sheet and $p$ in perpendicular direction. The dispersion relation is as usual,
\eq{1.22}{\om^2&=k^2+p^2.
}
The related by \Ref{1.11} phase shifts are
\eq{1.23}{\delta_{\rm TE}=-\arctan\frac{\Om}{p}, \ \
                \delta_{\rm TM}=-\frac{\pi}{2}+\arctan\frac{\om^2}{\Om p}.
}
with $\Om$ is defined in \Ref{3.1}. We mention that the phases \Ref{1.23} are written in a way that the arctangens do not leave the interval $[\frac{\pi}{2},\frac{\pi}{2}]$.

The spectrum of the \elm field consists of photonic (scattering) modes having real $\om$ and $p$ and, for the TM polarization, of surface modes having imaginary momentum, $p=i\eta$. The wave functions of these modes decrease exponentially in direction perpendicular to the sheet. The frequency $\om_{\rm sf}(k)$ of this mode is determined by the pole of the transmission coefficient, i.e., it is solution of the equation $1+Q_{\rm TM}=0$. With \Ref{1.21}   the  equation can be written in the form
\eq{1.24}{\om &= \frac{\Om}{\sqrt{2}}\sqrt{\sqrt{1+\left(\frac{k}{\Om/{2}}\right)^2}-1},
}
where for $\om_0\ne 0$ the frequency $\om$ is implicit in the right side by virtue of \Ref{3.1}.
We mention that the role of the surface modes in the Casimir effect was investigated in \cite{intr05-94-110404} and in \cite{bord05-39-6173}.
%
\subsubsection{\label{T2.2.2}Dielectric slab with plasma model}
In this model we consider a slab of finite thickness $L$ and permittivity given by the so-called 'plasma model',
\eq{1.25}{\ep(\om)&=1-\frac{{\om_p}^2}{\om^2+i0},
}
where ${\om_p}$ is the plasma frequency of the model. Also in this case a electron fluid is considered, now three dimensional, filling the slab, with the usual coupling to the \elm field. In this case the Maxwell equations result in dispersion relations
\eq{1.26}{\om^2 &=k^2+p^2, \ \ \mbox{outside the slab},
\nn\\   \ep(\om)\om^2 &=k^2+q^2, \ \ \mbox{inside the slab},
}
and the well known matching conditions on the surfaces of the slab. The corresponding transmission coefficients read
\eq{1.27}{ t_{\rm TE} &=\frac{4 p q e^{-i p L}}
                {(p+q)^2e^{-iqL}-(p-q)^2e^{iqL}},
\nn\\       t_{\rm TM} &=\frac{4\ep(\om) p q e^{- ip L}}
                {(\ep(\om)p+q)^2e^{-iqL}-(\ep(\om)p-q)^2e^{iqL}}.
}
The corresponding scattering phase shifts can be derived using \Ref{1.11}. However, these are not very useful for the numerical calculations in section \ref{T4} and we do not write them down here.

Also in this model there are surface plasmons in the TM polarization, having now both imaginary, $p=i\eta$ and $q=i\ga$
.  Writing the dispersion relation \Ref{1.26} with permittivity \Ref{1.25} in the form
\eq{1.28}{\om^2 &={\om_p}^2+k^2+q^2,
}
we see that  with imaginary $q$ and $p$ still one can have real frequency $\om$. The frequency $\om_{\rm sf}(k)$ of the plasmon is as before defined by the pole of the transmission coefficient and it is solution of the transcendent equation
\eq{1.29}{\coth(\ga L)&=\frac{\ep(\om)^2\eta^2+\ga^2}{2\ep(\om)\eta\ga}.
}
We mention the solution for large $L$, which is at once the plasmon travelling on a single surface,
\eq{1.30}{\om_{\rm sf}^{\rm single}(k)=\frac{{\om_p}}{\sqrt{2}}
        \sqrt{1+\left(\frac{k}{{\om_p}/\sqrt{2}}\right)^2
        -\sqrt{\left(\frac{k}{{\om_p}/\sqrt{2}}\right)^4+1}},
}
and we mention the bound
\eq{1.31}{\om_{\rm sf}(k)&\le \frac{{\om_p}}{\sqrt{2}}.
}
For a more detailed discussion we refer to the end of sect. II in \cite{bord12-85-025005}, where also the wave guide modes were discussed which are, however, not present in the  geometry used in the present paper.

\section{\label{T4}Free energy and entropy}
In this section we calculate the temperature dependent part of the free energy \Ref{1.10} and the entropy \Ref{1.11a}.
\subsection{\label{T4.1}Plasma model}

We start with the plasma model and consider eq. \Ref{1.10} for the free energy, whereby we disregard for the moment the contribution from the surface plasmon. We need the derivatives of the phase shifts \Ref{1.23} with account for \Ref{3.1},
\eq{3.1.1}{\frac{\pa}{\pa p}{\delta_{\rm TE}}&=
    \frac{\Om_0(k^4-k^2(\om_0^2-2p^2)+p^2(\om_0^2+p^2))}
    {k^4\om_0^2+((k^2-\om_0^2)^2+2k^2\Om_0^2)p^2+(2k^2-2\om_0^2+\Om_0^2)p^4+p^6},
\nn\\   \frac{\pa}{\pa p}{\delta_{\rm TM}}&=
    \frac{\Om_0(\om_0^2+p^2-k^2)}{(k^2-\om_0^2)^2+(2k^2-2\om_0^2+\Om_0^2)p^2+p^4}.
}
Here, with \Ref{3.1},  $\Im(\om)>0$ is assumed.

In the integrations over $\bm{k}$ and $p$, which form a half $\mathbb{R}^3$,   it is meaningful to change for spherical coordinates $p=\ep \om$, $k=\sqrt{1-\ep^2}\,\om$, where we used $\ep=\cos(\theta)\in [0,1]$. We get
\eq{4.2}{\Delta_TF_{\rm TX}&=\frac{T}{2\pi^2}\int_0^\infty d\om \,\om^2
\ln\left(1-e^{-\om/T}\right)h_{\rm TX}(\om)
}
with
\eq{4.3}{h_{\rm TX}(\om)=\int_0^1 d\ep\, \delta_{\rm TX}'(\om).
}
For the two polarizations we insert the corresponding expressions. The integrations can be carried out explicitly,
\eq{4.4}{h_{\rm TE}(\om) &=
    \frac{2\om(\om^2-\om_0^2)\om_0^2\Om_0+((\om^2-\om_0^2)^2-2\om^2\om_0^2\Om_0^2)
    \,{\rm arccot}\left(\frac{\om\Om_0}{\om^2-\om_0^2}\right)}
    {\om(\om^2-\om_0^2)^2},
\nn\\   h_{\rm TM}(\om) &=
    \frac{2\om \Om_0-(2\om^2-2\om_0^2+\Om_0^2)
    \,{\rm arccot}\left(\frac{\om\Om_0}{\om^2-\om_0^2}\right)}
    {\om\Om_0^2} .
}
Inserting \Ref{4.4} into \Ref{1.10},  the tempera\-ture dependent part of the free energy appears represented by a single integration. We consider its behavior at high temperature. For this to do we make the substitution $\om\to\om T$,
\eq{4.5}{\Delta_T{\cal F} &=\frac{T^4}{2\pi^2} \int_0^\infty d\om \,\om^2 \ln\left(1-e^{-\om}\right)h_{\rm TX}(\om T).
}
Using the expansions for large argument,
\eq{4.6}{h_{\rm TE}(\om )=\frac{\pi}{2\om }-\frac{\Om_0}{\om^2}\dots,
\ \ h_{\rm TM}(\om )=-\frac{\Om_0}{3\om^2}+\dots,
}
we obtain for $T\to\infty$
\eq{4.7}{\Delta_T{\cal F}_{\rm TE} &= -\frac{T^3\zeta(3)}{4\pi}+\frac{\Om_0  T^2}{12}+O(T\ln(T)),
\nn\\   \Delta_T{\cal F}_{\rm TM} &= \frac{\Om_0  T^2}{36}+O(T\ln(T)).
}
These are contributions growing faster than the first power in temperature and  must be subtracted according to the discussion in Sect. 2. That can be achieved by the following subtractions in $h_{\rm TX}$,
\eq{4.8}{h_{\rm TE}^{\rm subtr.}(\om)&=
 h_{\rm TE} (\om)-\frac{\pi}{2\om}+\frac{\Om_0}{\om^2},
\nn\\   h_{\rm TM}^{\rm subtr.}(\om)&=
h_{\rm TM} (\om) +\frac{\Om_0}{3\om^2}.
}
With these functions, the subtracted free energy becomes
\eq{4.9}{\Delta_TF_{\rm TX}^{\rm subtr.}&=
\frac{T}{2\pi^2}\int_0^\infty d\om \,\om^2
\ln\left(1-e^{-\om/T}\right)h_{\rm TX}^{\rm subtr.}(\om)
}
and the subtracted entropy becomes
\eq{4.10}{S^{\rm subtr.}_{\rm TX}&=
\frac{1}{2\pi^2}\int_0^\infty d\om \,\om^2g\left(\frac{\om}{T}\right)h_{\rm TX}^{\rm subtr.}(\om)
}
with the function $g(x)$ defined in \Ref{1.11b}.
%
Further we have to consider the contribution from the surface plasmon.
The solution of the defining equation \Ref{1.24} is
\eq{4.sf}{\om_{\rm sf}(k) &= \sqrt{\om_0^2-\frac{1}{2}\Om_0^2+\Om_0\sqrt{k^2-\om_0^2+\frac14 \Om_0^2}}
}
and it is real for $\om_0>0$. Its contribution to the free energy reads
\eq{4.12}{\Delta_T{\cal F}_{\rm sf}=\frac{T}{2\pi}\int_{\om_0}^\infty dk\,k  \ln\left(1-e^{-\om_{\rm sf}(k)/T}\right).
}
Changing variables from ${k}$ to $\om$ we arrive at
\eq{4.13}{\Delta_T{\cal F}_{\rm sf}=
\frac{T}{2\pi}\int_{\max(0,\sqrt{\om_0^2-\Om_0^2/2})}^\infty d\om \,\om\left(1-\frac{2\om_0^2}{\Om_0^2}+\frac{2\om^2}{ \Om_0^2}\right)
    \ln\left(1-e^{-\om/T}\right).
}
For $\om_0>\Om_0/\sqrt{2}$ the integration can be rewritten as $\int_0^\infty -\int_0^{\sqrt{\om_0^2-\Om_0^2/2}}$.
The integration in the first integral can be carried out explicitly,
\eq{4.14}{\Delta_T{\cal F}_{\rm sf}&= -\left(1-\frac{2\om_0^2}{\Om_0^2}\right)\frac{T^3\zeta(3)}{2\pi}-\frac{6T^5\zeta(5)}{ \pi\Om_0^2}
\\\nn &~~~
-\Theta\left(\om_0-\frac{\Om_0}{\sqrt{2}}\right)
\frac{T}{2\pi}\int^{\sqrt{\om_0^2-\Om_0^2/2}}_0 d\om \,\om\left(1-\frac{2\om_0^2}{\Om_0^2}+\frac{2\om^2}{ \Om_0^2}\right)
    \ln\left(1-e^{-\om/T}\right),
}
where the step function indicates that the second contribution is present for $\om_0>\Om_0/2$ only.

The first two contributions, which are powers of $T$, must be subtracted in accordance with \Ref{1.20} and from \Ref{4.14} the integral contribution,
\eq{4.14a}{\Delta_T{\cal F}_{\rm sf}^{\rm subtr.}&= -\Theta\left(\om_0-\frac{\Om_0}{\sqrt{2}}\right)
\frac{T}{2\pi}\int^{\sqrt{\om_0^2-\Om_0^2/2}}_0 d\om \,\om\left(1-\frac{2\om_0^2}{\Om_0^2}+\frac{2\om^2}{ \Om_0^2}\right)
    \ln\left(1-e^{-\om/T}\right),
}
remains.
With \Ref{1.11a}  we get accordingly for the entropy
\eq{4.15}{ S^{\rm subtr.}_{\rm sf}&= -\Theta\left(\om_0-\frac{\Om_0}{\sqrt{2}}\right)
\frac{1}{2\pi}\int^{\sqrt{\om_0^2-\Om_0^2/2}}_0 d\om \,\om\left(1-\frac{2\om_0^2}{\Om_0^2}+\frac{2\om^2}{ \Om_0^2}\right)
g\left(\frac{\om}{T}\right).
}

This way, the complete temperature dependent part of the free energy, following from \Ref{4.9} and \Ref{4.12}, and that of the entropy, following from \Ref{4.10} and \Ref{4.15}, are
\eq{4.F}{\Delta_T{\cal F}^{\rm subtr} &=
\Delta_T{\cal F}^{\rm subtr}_{\rm TE}+\Delta_T{\cal F}^{\rm subtr}_{\rm TM}+\Delta_T{\cal F}^{\rm subtr}_{\rm sf},
\nn\\  S^{\rm subtr} &=
S^{\rm subtr}_{\rm TE}+S^{\rm subtr}_{\rm TM}+S^{\rm subtr}_{\rm sf}.
}
These are the final expressions for the temperature dependent part of the free energy and the entropy. Obviously, the entropy is a dimensionless function of $\Om_0 T$ and $\om_0 T$. Since it is  represented by single, fast converging integration, the  numerical evaluation is straightforward. Plots are shown in Fig. 1.

\begin{figure}\label{fig1}
    \includegraphics[width={0.5\textwidth}]{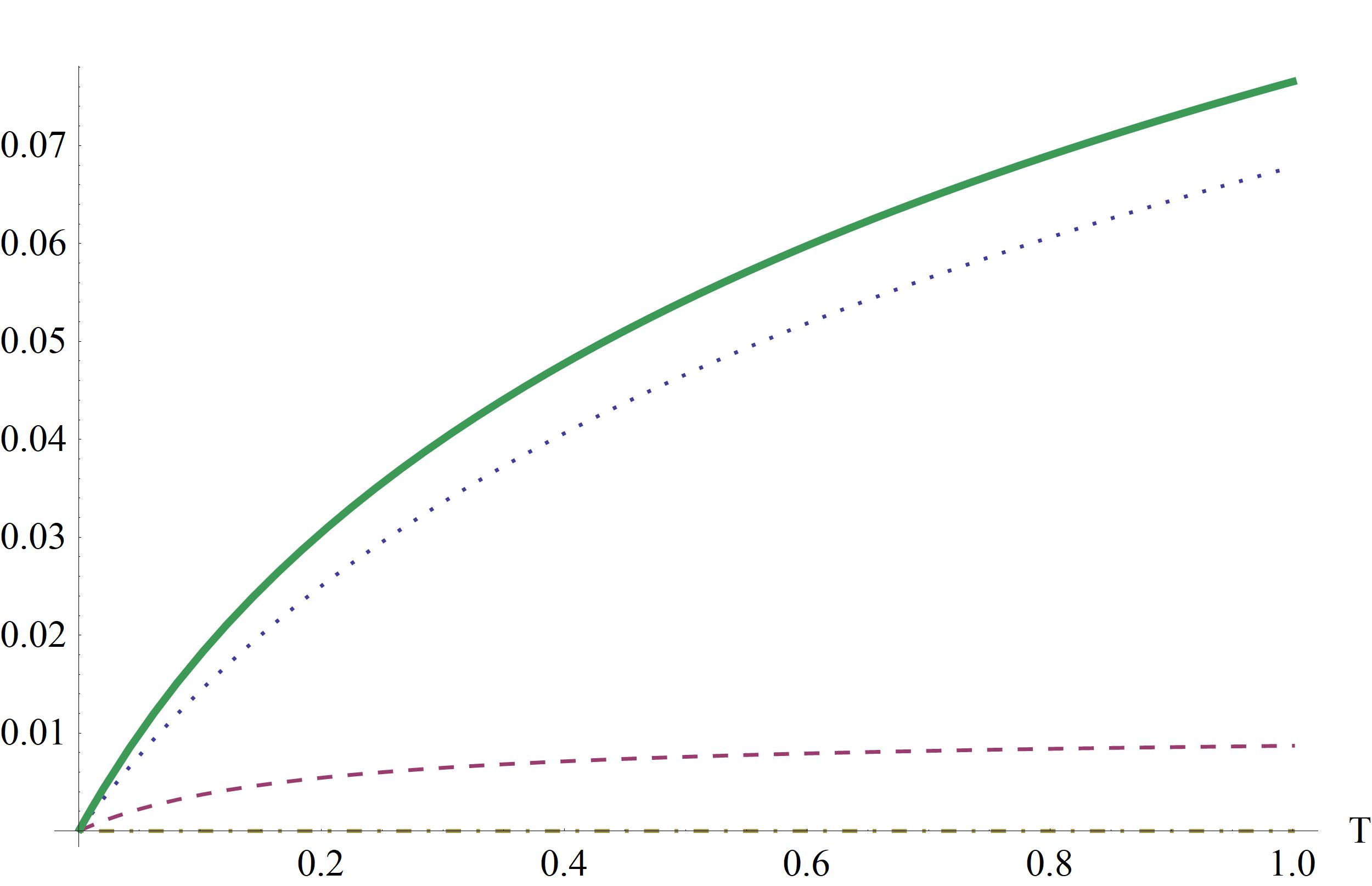}
 \ \    \includegraphics[width={0.5\textwidth}]{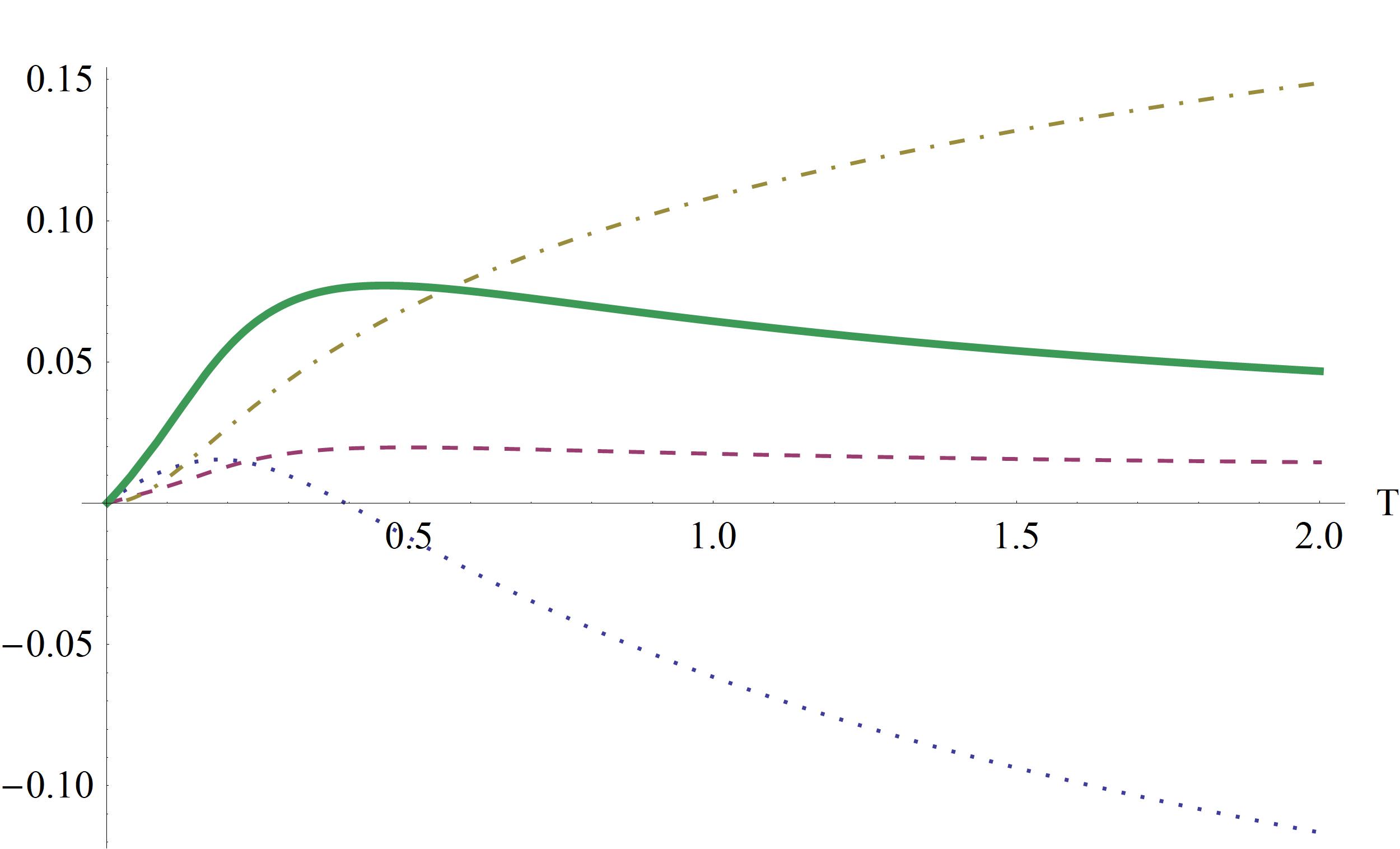}
    \caption{The entropy \Ref{4.F} of a flat plasma sheet (solid line)  with $\om_0=0$ (left panel) and $\om_0=1.16\Om_0$ (right panel). The dotted line is the TE-contribution, the dashed   line is the TM-contribution and the dot-dashed line is the contribution from the surface plasmon.
    }
\end{figure}

The expansion for small temperature can be obtained from \Ref{4.10} by inserting the expansions of the functions $h_{\rm TX}^{\rm subtr.}(\om)$ for small argument. The leading order, however, follows from the subtraction terms and is
\eq{4.11}{S^{\rm subtr.}_{\rm TE}&=\frac{\Om_0 T}{6}+O(T^2),
\ \   S^{\rm subtr.}_{\rm TM}=\frac{\Om_0 T}{18}+O(T^2).
}
From the surface plasmon we have with \Ref{4.14} $S^{\rm subtr.}_{\rm sf} = O(T)$, demonstrating together with \Ref{4.11} that  Nernst's theorem is satisfied for this model.

Next to discuss is the behavior of the entropy for high temperatures.
For the free energy we expand the logarithm in \Ref{4.9} and get
\eq{4.16}{ {\cal F}_{\rm TX}^{\rm subtr.} &= \frac{1}{2\pi^2}\int_0^\infty d\om\,\om^2\,
h_{\rm TX}^{\rm subtr.}(\om)\ T\ln(T)+O(T).
}
The integral is converging due to the decrease of the function $h_{\rm TX}^{\rm subtr.}(\om)$ for $\om\to\infty$ which results from the subtractions done in \Ref{4.8}.
From \Ref{4.14a} we get in a similar way
\eq{4.17}{\Delta_T{\cal F}_{\rm sf}^{\rm subtr.}&=
            -\frac{1}{4\pi\Om_0^2}\left(\om_0^2-\frac{\Om_0^2}{2}\right)^2\ T\ln(T)+O(T)
}
from the surface plasmon's contribution.

For the entropy we expand the function $g$ in \Ref{4.10},
\eq{4.g}{g\left(\frac{\om}{T}\right)=\ln(T)+\dots
}
in \Ref{4.10} and \Ref{4.15} and arrive at
\eq{4.18}{S^{\rm subtr.}_{\rm TX}&=
\frac{1}{2\pi^2}\int_0^\infty d\om \,\om^2 h_{\rm TX}^{\rm subtr.}(\om)\ \ln(T)+O(1),
\nn\\
         S^{\rm subtr.}_{\rm sf} &=
         \frac{\left(\om_0^2-\frac{\Om_0^2}{2}\right)^2}{4\pi \Om_0^2}\ \ln(T)+O(1).
}
From the above asymptotic expansions for $T\to\infty$  one can get the heat kernel coefficients from comp[aring with the general expansion  \Ref{1.18}. Using \Ref{4.7}, \Ref{4.16} and \Ref{4.17} we get
\eq{4.19}{  a_{\frac12}^{\rm TE} &=\sqrt{\pi},  &  a_{1}^{\rm TE} &= -2\Om_0,
\nn \\  a_{\frac12}^{\rm TM} &=2\sqrt{\pi}\left(1-2\frac{\om_0^2}{\Om_0^2}\right),
  &  a_{1}^{\rm TM} &= -\frac23\Om_0.
}
From \Ref{4.16} and \Ref{4.17} we get
\eq{4.20}{ a_{\frac32}^{\rm TE} &=
\frac{4}{\sqrt{\pi}} \int_0^\infty d\om\,\om^2\,h_{\rm TE}^{\rm subtr.}(\om),
\ \   a_{\frac32}^{\rm TM} &=
 \frac{2\sqrt{\pi}}{\Om_0^2} \left(\om_0^2-\frac{\Om_0^2}{2}\right)^2\Theta\left(\om_0-\frac{\Om_0}{\sqrt{2}}\right).
}
For $\om_0=0$ we get back the coefficients derived in \cite{bord05-38-11027} (in eq. (4.29) it must be $B_{1/2}=2\sqrt{\pi}$). As mentioned in \cite{bord05-38-11027}, the heat kernel $K(t)$ has a nonstandard behavior for $t\to0$ which corresponds to the $T^5$-term in \Ref{4.14} and can be formally expressed as heat kernel coefficient with negative number, $a_{-1/2}=4\sqrt{\pi}\Om_0^{-2}$.

It is interesting to mention that in \Ref{4.16} for the TM case $\int_0^\infty d\om\,\om^2\,h_{\rm TM}^{\rm subtr.}(\om)=0$ holds. The behavior of $a_{\frac32}^{\rm TE}$ as function of the intrinsic oscillator frequency $\om_0$ is shown in Fig. 3. As can be seen, it changes sign at $\om_0=\Om_0/\sqrt{2}$. For larger $\om_0$ it becomes negative and competes with $a_{\frac32}^{\rm TM}$, \Ref{4.20}, which results from the surface plasmon. There is a small region where their sum is negative which makes the entropy for large temperature taking negative values as shown in Fig. 3.

\begin{figure}\label{fig2}
\begin{minipage}{0.48\textwidth}
    \includegraphics[width={0.98\textwidth}]{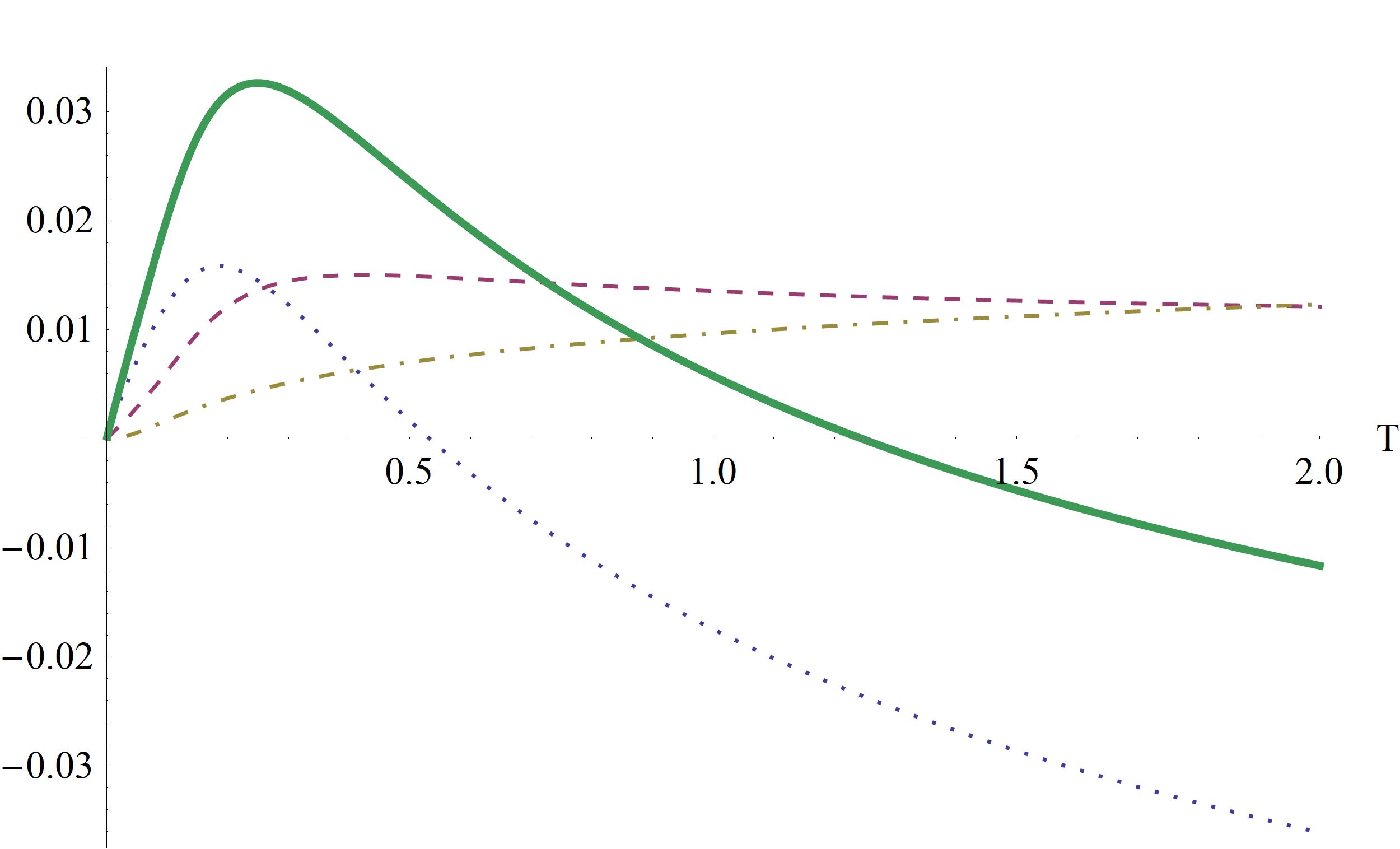}
    \caption{The entropy \Ref{4.F} of a flat plasma sheet (solid line)  with $\om_0=0.85\Om_0$. The dotted line is the TE-contribution, the dashed   line is the TM-contribution and the dot-dashed line is the contribution from the surface plasmon.
    }
\end{minipage} \ \
\begin{minipage}{0.48\textwidth}
    \includegraphics[width={1\textwidth}]{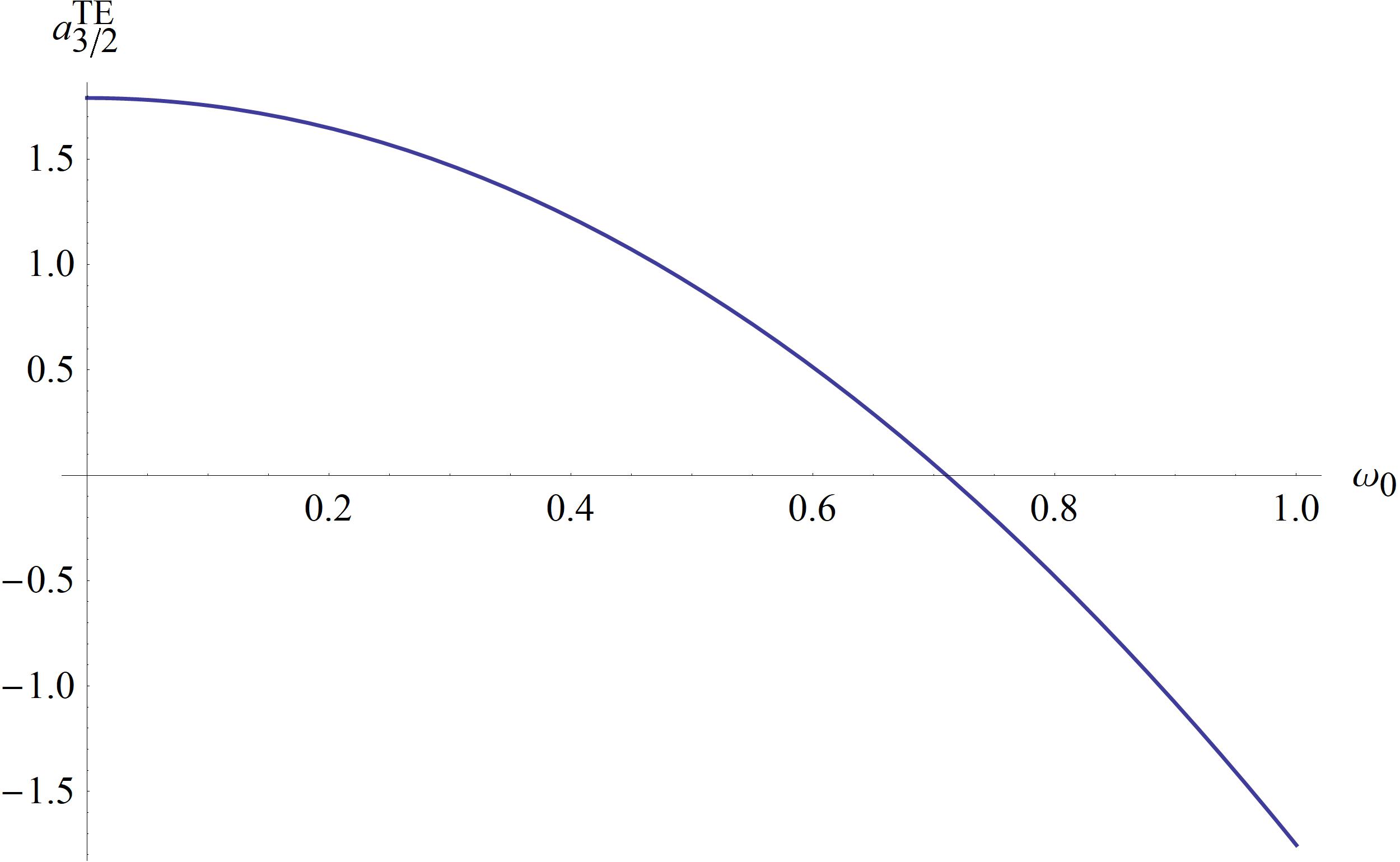}
    \caption{The heat kernel coefficient $a^{\rm TE}_{3/2}$, eq. \Ref{4.20}, as function of $\om_0$. It changes sign at $\om_0=\frac{\Om_0}{2}$. }
\end{minipage}\ \ \
\end{figure}

%
\subsection{\label{T4.2}Dielectric slab}
In this subsection we consider the dielectric slab as defined in sect. \ref{T2.2.2}.
We represent the transmission coefficients \Ref{1.27} as products
\eq{3.2.1}{ t_{\rm TX}&=t_{\rm TX}^{\rm s}  \,  t_{\rm TX}^{\rm L}\, e^{i(q-p)L},
}
where
\eq{3.2.2}{ t_{\rm TE}^{\rm s}  &=  \frac{4pq}{(p+q)^2},\ \
             t_{\rm TM}^{\rm s}  =  \frac{4\ep(\om)pq}{(\ep(\om)p+q)^2}
}
are the contributions from the surfaces of the slab,
\eq{3.2.3}{ t_{\rm TE}^{\rm L}  &=  \frac{1}{1-\left(\frac{p+q}{p+q}\right)^2e^{2iqL}},\ \
             t_{\rm TM}^{\rm L}  = \frac{1}{1-\left(\frac{\ep(\om)p+q}{\ep(\om)p+q}\right)^2e^{2iqL}},
}
are the contributions depending on the thickness $L$ of the slab, i.e., on the thickness between the two surfaces, and the exponential in \Ref{3.2.1} is what remains. As we will see below the latter  gives a contribution which is proportional to the thickness of the slab.
According to \Ref{1.11}, the factorization \Ref{3.2.1} delivers a sum of the corresponding phase shifts,
\eq{3.2.4}{\delta_{\rm TX}(p) &= \delta_{\rm TX}^{\rm s}(p)+
\delta_{\rm TX}^{\rm L}(p)+(p-q)L ,
}
and by means of \Ref{1.10} of the free energy,
\eq{3.2.5}{{\cal F}_{\rm TX}(p) &= {\cal F}_{\rm TX}^{\rm s} +{\cal F}_{\rm TX}^{\rm L} +  {\cal F}^{\rm exp}.
}
We consider these contributions separately.
%
\subsubsection{\label{T4.2.1}Thickness independent contribution ${\cal F}_{\rm TX}^{\rm s}$}

We start with the thickness independent contributions from the surfaces \Ref{3.2.2}. When inserting into \Ref{1.11} we have a nonzero contributions only from the momentum region $p<{\om_p}$, where $q$ is imaginary, $q=i\ga$ with $\ga=\sqrt{{\om_p}^2-p^2}$. The contribution to the free energy is
\eq{3.2.6}{{\cal F}_{\rm TX}^{\rm s}&=  \int\frac{d\bm{k}}{(2\pi)}
\int_0^{\om_p}\frac{dp}{\pi} T\ln\left(1-e^{-\om/T}\right)\frac{\pa}{\pa p}
\delta_{\rm TX}^{\rm s}(p).
}
We continue with the TE contribution. With \Ref{3.2.2} we get
\eq{3.2.7}{\delta_{\rm TE}^{\rm s}(p)&=\frac{\pi}{2}-2\arctan\left(\frac{\ga}{p}\right).
}
We change the integration in \Ref{3.2.6} from $k$ to $\om=\sqrt{k^2+p^2}$,
\eq{3.2.8}{{\cal F}_{\rm TE}^{\rm s}&=
\frac{T}{2\pi^2}\int_0^\infty d\om\,\om \ln\left(1-e^{-\om/T}\right)
\int_0^{\min(\om,{\om_p})} dp  \frac{\pa}{\pa p} \delta_{\rm TE}^{\rm s}(p).
}
Since $\delta_{\rm TE}^{\rm s}(p)$ does not depend on $\om$ (in opposite to $\delta_{\rm TM}^{\rm s}(p)$ which does through $\ep(\om)$), the integration over $p$ can be carried out,
\eq{3.2.9}{{\cal F}_{\rm TE}^{\rm s}&=
\frac{T}{2\pi^2}\int_0^{\om_p} d\om\,\om \ln\left(1-e^{-\om/T}\right)
\left( \delta_{\rm TE}^{\rm s}(\om)+\frac{\pi}{2}\right)
+\frac{T}{2\pi}\int_{\om_p}^\infty d\om\,\om \ln\left(1-e^{-\om/T}\right),
}
where we accounted for $\delta_{\rm TE}^{\rm s}(0)=-\frac{\pi}{2}$ and  $\delta_{\rm TE}^{\rm s}({\om_p})=\frac{\pi}{2}$. We rearrange the integrations,
\eq{3.2.10}{{\cal F}_{\rm TE}^{\rm s}&=
\frac{T}{2\pi}\int_0^\infty d\om\,\om \ln\left(1-e^{-\om/T}\right)
+\frac{T}{2\pi^2}\int_{\om_p}^\infty d\om\,\om \ln\left(1-e^{-\om/T}\right)
\left( \delta_{\rm TE}^{\rm s}(\om)-\frac{\pi}{2}\right).
}
The first integration is explicit and with \Ref{3.2.7} we arrive at
\eq{3.2.11}{{\cal F}_{\rm TE}^{\rm s}&=
-\frac{\zeta(3)}{2\pi}T^3-
\frac{T}{\pi^2}\int_0^{\om_p} d\om\,\om \ln\left(1-e^{-\om/T}\right)
\arctan\left(\frac{\ga}{\om}\right).
}
We mention that for $T\to0$ the $T^3$-contributions cancel and the expression starts with $T^4$ as can be seen easily. For $T\to\infty$  we expand the logarithm and get
\eq{3.2.12}{{\cal F}_{\rm TE}^{\rm s}&=
-\frac{\zeta(3)}{2\pi}T^3-
\frac{T}{\pi^2}\int_0^{\om_p} d\om\,\om \ln\left(\frac{\om}{T}\right)
\arctan\left(\frac{\ga}{\om}\right)+\dots\,.
}
The integration can be carried out and the behavior for $T\to\infty$ is
\eq{3.2.13}{  {{\cal F}_{\rm TE}^{\rm s}}^{\rm subtr.}&=
\frac{{\om_p}^2}{8\pi}T\ln\left(\frac{2T}{{\om_p}}\right)+\dots\,,
}
where we subtracted the $T^3$-term according to \Ref{1.20}.

Now we consider the TM-contribution. Here it is not possible to integrate simply the derivative like in \Ref{3.2.8} since the phase shift depends on the frequency. Nevertheless, this integration can be done explicitly.
With \Ref{3.2.2} and \Ref{1.11} we get
\eq{3.2.14}{\delta_{\rm TM}^{\rm s}(p,\om)&=
-\frac{\pi}{2}+2\arctan\left(\left(1-\frac{{\om_p}^2}{\om^2}\right)
\frac{p}{\sqrt{{\om_p}^2-p^2}}\right)
}
for $p\le{\om_p}$, otherwise it is zero. We mention the special cases
\eq{3.2.15}{\delta_{\rm TM}^{\rm s}(0,\om)&=-\frac{\pi}{2}, \ \ \
\delta_{\rm TM}^{\rm s}({\om_p},\om)=\left\{\begin{array}{rl}
-\frac{3\pi}{2},&(\om<{\om_p}),\\ \frac{\pi}{2},&(\om >{\om_p}).
\end{array}\right.
}
Next we consider the free energy \Ref{3.2.6} and integrate by parts. We get
\eq{3.2.16}{{\cal F}_{\rm TM}^{\rm s}&= A+B,
}
where we introduced the notations
\eq{3.2.17}{A &= \int\frac{d\bm{k}}{(2\pi)^2}
\frac{T}{\pi}   \left(  \ln\left(1-e^{-\sqrt{k^2+\om_p^2}/T}\right)
 \delta_{\rm TM}^{\rm s}({\om_p},\om)
 -   \ln\left(1-e^{-k/T}\right)   \delta_{\rm TM}^{\rm s}(0,\om)\right),
\nn\\  B   &=\int\frac{d\bm{k}}{(2\pi)^2}\int_0^{\om_p}\frac{dp}{\pi}
\frac{p}{\om}\frac{-1}{e^{\om/T}-1}
\delta_{\rm TM}^{\rm s}(p,\om).
}
Here, $A$ denotes the surface term. Using \Ref{3.2.15} it simplifies and with a change of the integration variable it can be written in the form
\eq{3.2.18}{A &= \frac{T}{4\pi}\left(
    2\int_0^\infty d\om\,\om   \ln\left(1-e^{-\om/T}\right)
     - \int_0^{\om_p} d\om\,\om   \ln\left(1-e^{-\om/T}\right) \right),
\nn\\  &= -\frac{\zeta(3)}{2\pi}T^3-
\frac{T}{4\pi}\int_0^{\om_p} d\om\,\om   \ln\left(1-e^{-\om/T}\right) .
}
The behavior for $T\to0$ has contributions from both terms in the last line and reads
\eq{3.2.19}{A &=-\frac{\zeta(3)}{4\pi}T^3+\dots\,.
}
For $T\to\infty$ we expand the logarithm and get
\eq{3.2.20}{A &= -\frac{\zeta(3)}{2\pi}T^3
    +\frac{{\om_p}^2 T}{8\pi}\left(\ln\left(\frac{T}{{\om_p}}\right)+\frac12\right)+\dots\,.
}
Now we consider $B$, \Ref{3.2.17}, and make a substitution of variables $k\to \om=\sqrt{k^2+p^2}$,
\eq{3.2.21}{ B &= -\frac{1}{2\pi^2}\int_0^\infty \frac{d\om\,\om}{e^{\om/T}-1}h(\om),
}
where
\eq{3.2.22}{h(\om) &= \int_0^{\min(\om,{\om_p})} dp\,
            \delta_{\rm TM}^{\rm s}(p,\om)
\equiv \left\{\begin{array}{rl}
h_1(\om),&(\om<{\om_p}),\\ h_2(\om),&(\om >{\om_p}).
\end{array}\right.
}
These functions can be calculated explicitly,
\eq{3.2.23}{h_1(\om) &=
    -\frac{\pi\om}{2}
    +\frac{2\om}{\sqrt{{\om_p}^2-2\om^2}}
    \left[ -\sqrt{{\om_p}^2-2\om^2}\,{\rm arccot}
    \left(\frac{\om}{\sqrt{{\om_p}^2-\om^2}}\right)
\right.
\nn\\\nn & \ \ \ \ +\left.
\om \ {\rm arccoth}\left(\frac{\om^2-{\om_p}^2}{\sqrt{2\om^4-3\om^2{\om_p}^2+{\om_p}^4}}\right)
    +\om\
    {\rm arctanh}\left(\frac{{\om_p}\sqrt{{\om_p}^2-2\om^2}}{{\om_p}^2-\om^2}\right)\right],
\\ h_2(\om)&= \frac{\pi{\om_p}}{2}+\frac{2\om^2}{\sqrt{2\om^2-{\om_p}^2}}
{\rm arctan}\left(\frac{{\om_p}\sqrt{2\om^2-{\om_p}^2}}{{{\om_p}^2-\om^2}}\right).
}
Their asymptotics read
\eq{3.2.24}{ h(\om) &\raisebox{-4pt}{$=\atop \om\to0$}  -\frac{3\pi}{2}\om+\dots\,
, \ \ \   h(\om) \raisebox{-4pt}{$=\atop \om\to\infty$}  -\frac{\pi-4}{2}{\om_p}-\frac{2{\om_p}^3}{3\om^2}+\dots\,.
}
With the substitution $\om\to\om T$ in \Ref{3.2.21} we find immediately for $T\to0$
\eq{3.2.25}{ B&= \frac{3\zeta(3)}{2\pi}T^3+\dots\,.
}
In order to get the behavior for $T\to\infty$ we rewrite the integral \Ref{3.2.21} in the form
\eq{3.2.26}{B &= -\frac{1}{2\pi^2}\int_0^\infty \frac{d\om\,\om}{e^{\om/T}-1}\frac{\pi-4}{2}{\om_p}
 -\frac{1}{2\pi^2}\int_0^\infty \frac{d\om\,\om}{e^{\om/T}-1}\left(h(\om)-\frac{\pi-4}{2}{\om_p}\right).
}
The first integral is explicit and the parenthesis in the second integral decreases like $1/\om^2$ for $\om\to\infty$. That allows to expand the Boltzmann factor and we get
\eq{3.2.27a}{B &= \frac{4-\pi}{24}{\om_p} T^2+\frac{c}{2\pi}{\om_p}^2 T+\dots\,,
}
where $c=\int_0^\infty {d\om } \left(h(\om)_{|_{{\om_p}=1}}-\frac{\pi-4}{2} \right)\simeq 1.5708$.

Collecting from \Ref{3.2.19} and \Ref{3.2.25} we get for the free energy \Ref{3.2.16} for $T\to0$
\eq{3.2.26a}{{\cal F}_{\rm TM}^{\rm s}&=\frac{5\zeta(3)}{4\pi}T^3+\dots
}
and from \Ref{3.2.20} and \Ref{3.2.27} for $T\to\infty$
\eq{3.2.27}{{\cal F}_{\rm TM}^{\rm s}&=
-\frac{\zeta(3)}{2\pi}T^3+\frac{4\pi}{24}{\om_p} T^2
+\frac{{\om_p}^2 T}{8\pi}\left(\ln \left(\frac{T}{{\om_p}}\right)+\frac12+\frac{4c}{\pi}\right)+\dots\,.
}
Finally we have to define the subtracted free energy. Using \Ref{3.2.27} we get
\eq{3.2.28}{  {{\cal F}_{\rm TM}^{\rm s}}^{\rm subtr.}&=
    {{\cal F}_{\rm TM}^{\rm s}}+\frac{\zeta(3)}{2\pi}T^3-\frac{4\pi}{24}{\om_p} T^2.
}
By means of \Ref{1.11a}, taking the derivative with respect to temperature, we get also the subtracted entropy
${S_{\rm TE}^{\rm s}}^{\rm subtr.}$ and
${S_{\rm TM}^{\rm s}}^{\rm subtr.}$ as well as their sum. These are shown in Fig. 4.

\begin{figure}\label{fig4}
\includegraphics[width=7cm]{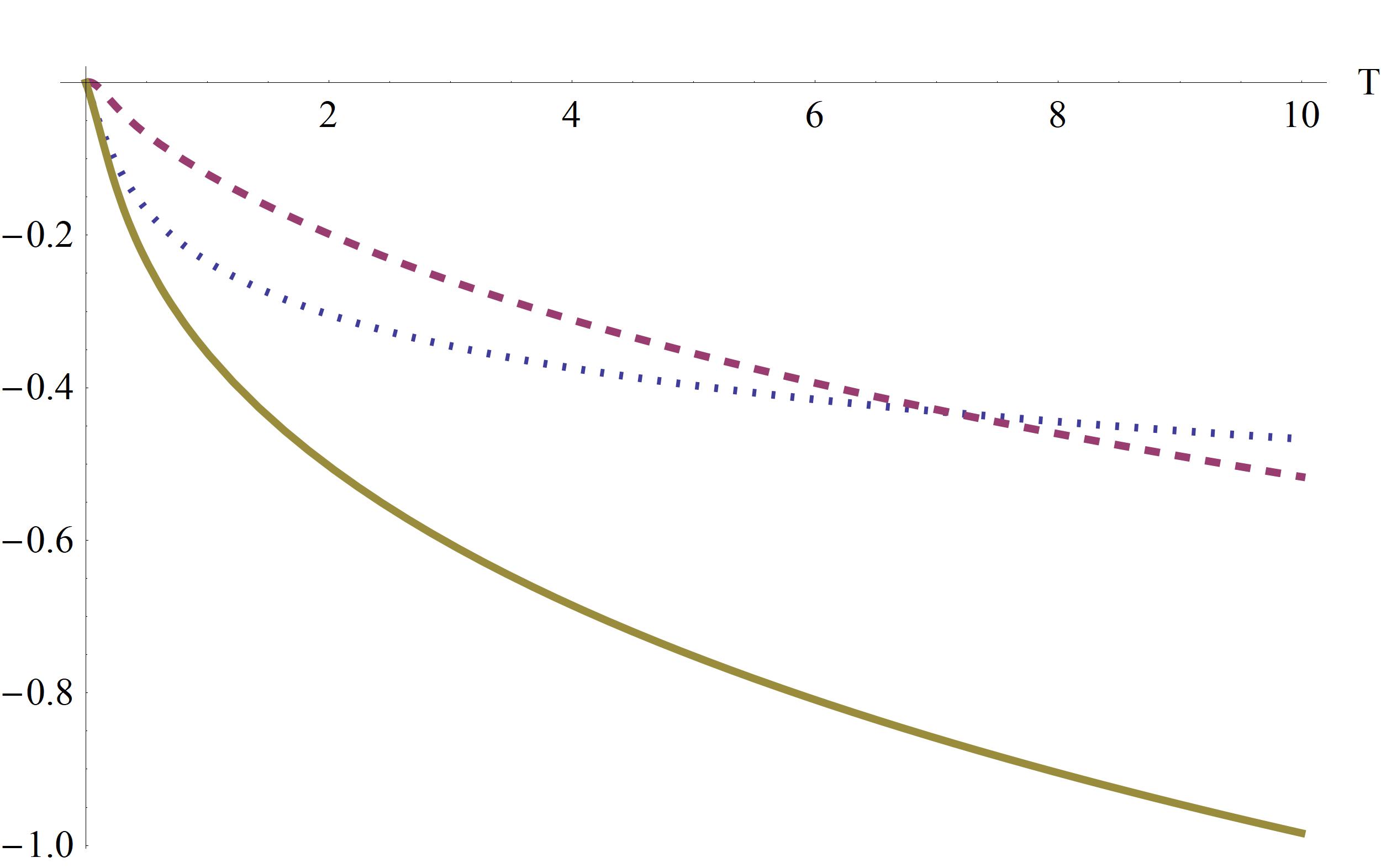}\ \ \ \ \includegraphics[width=7cm]{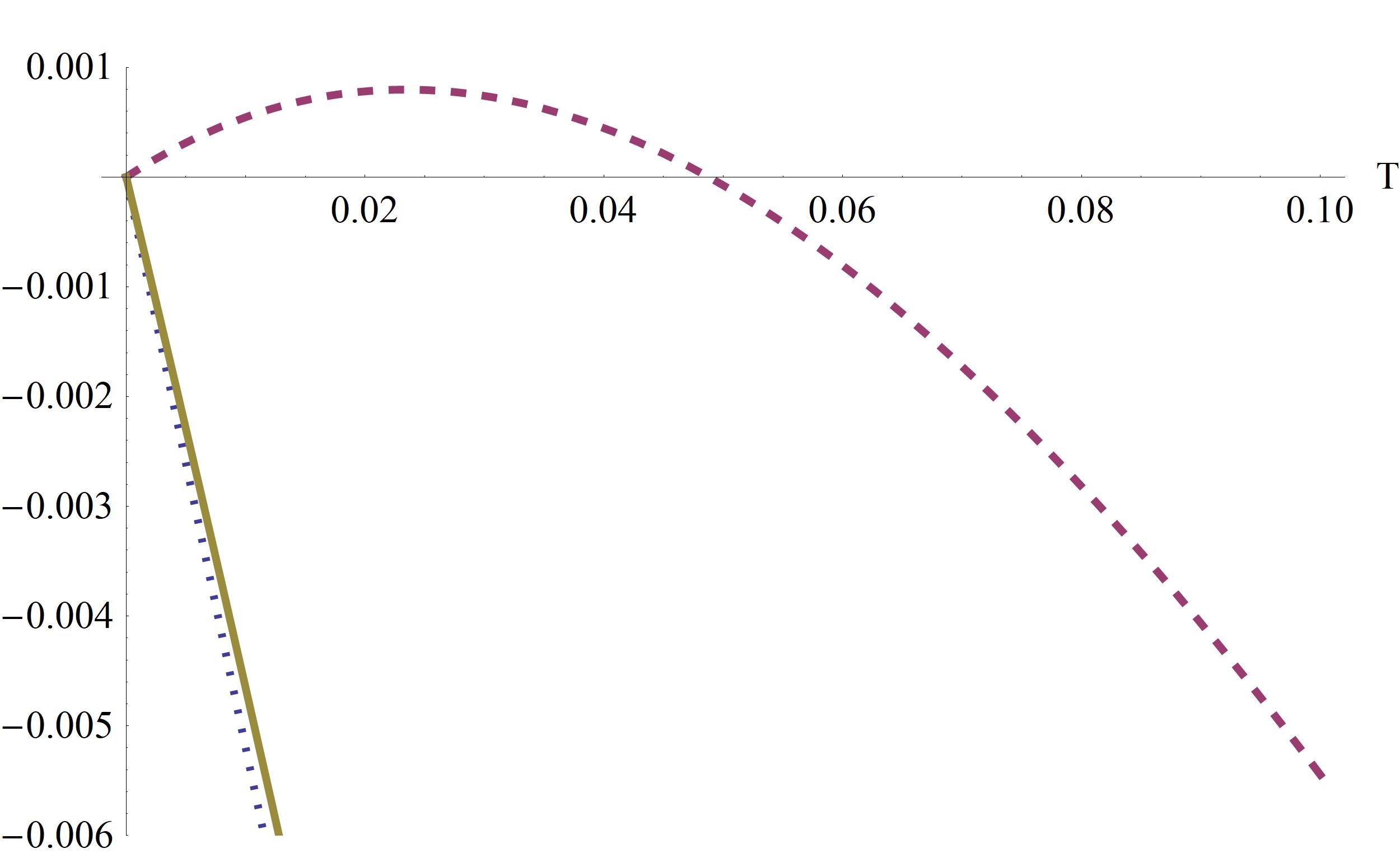}
\caption{The thickness independent part of the entropy \Ref{3.2.28} of a  dielectric slab with ${\om_p}=1$  as function of the temperature (left panel) and, separately,  the region of small temperature (right panel). The dotted line is the TE-contribution,
 the dashed   line is the TM-contribution and the solid line is the sum,
 ${S^{\rm s}}^{\rm ~subtr.}={S_{\rm TE}^{\rm s}}^{\rm subtr.}+{S_{\rm TM}^{\rm s}}^{\rm subtr.}$.}
\end{figure}
%
\subsubsection{\label{T4.2.2}Thickness dependent contribution ${\cal F}_{\rm TX}^{\rm L} $}

Now we consider the thickness (thickness) dependent part  which is given by the transmission coefficients $t^{\rm L}_{\rm TX}$, \Ref{3.2.3}. We call it {\it Lifshitz part} since it is just what one gets from the Lifshitz formula. Using \Ref{1.11} we get from \Ref{3.2.3} for $p<{\om_p}$
\eq{3.2.30}{ \delta^{\rm L}_{\rm TE}(p) &=
\frac{1}{2i}\ln\left(\frac{1-\left(\frac{p+i\ga}{p-i\ga}\right)^2 e^{-2\ga L}}
                                    {1-\left(\frac{p-i\ga}{p+i\ga}\right)^2 e^{-2\ga L}} \right),\ \
\delta^{\rm L}_{\rm TM}(p,\om) =
\frac{1}{2i}\ln\left(\frac{1-\left(\frac{\ep(\om)p+i\ga}{\ep(\om)p-i\ga}\right)^2 e^{-2\ga L}}
                                    {1-\left(\frac{\ep(\om)p-i\ga}{\ep(\om)p+i\ga}\right)^2 e^{-2\ga L}} \right)
}
with $\ga=\sqrt{{\om_p}^2-p^2}$ and for $p>{\om_p}$
\eq{3.2.31}{ \delta^{\rm L}_{\rm TE}(p) &=
\frac{1}{2i}\ln\left(\frac{1-\left(\frac{p-q}{p+q}\right)^2 e^{-2i q L}}
                                    {1-\left(\frac{p-q}{p+q}\right)^2 e^{2iq L}} \right),\ \
\delta^{\rm L}_{\rm TM}(p,\om) =
\frac{1}{2i}\ln\left(\frac{1-\left(\frac{\ep(\om)p-q}{\ep(\om)p+q}\right)^2 e^{-2iq L}}
                                    {1-\left(\frac{\ep(\om)p-q}{\ep(\om)p+q}\right)^2 e^{2iq L}} \right)
}
with $q=\sqrt{p^2-{\om_p}^2}$. We indicated explicitly that the phase shift for TE depends on $p$ only, whereas that for TM has also a dependence on  with $\om=\sqrt{k^2+p^2}$. The Lifshitz-part of free energy is given by the formula
\eq{3.2.32}{ {\cal F}^{\rm L}_{\rm TX} &=
    \int\frac{d\bm{k}}{(2\pi)^2}\int_0^\infty\frac{dp}{\pi}\,
    T\ln\left(1-e^{-\om/T}\right)\frac{\pa}{\pa p}
    \left\{\delta^{\rm L}_{\rm TE}(p),  \atop \delta^{\rm L}_{\rm TM}(p,\om).\right.
}
Again,  we consider first the TE-contribution. We change the integration over $k$ for $\om$ and since the phase shift does not depend on $\om$ we can carry out the integration over $p$,
\eq{3.2.33}{  {\cal F}^{\rm L}_{\rm TE} &=
    \frac{T}{2\pi^2}\int_0^\infty d\om\,\om\,\ln\left(1-e^{-\om/T}\right)
    \delta^{\rm L}_{\rm TE}(\om).
}
For the TM contribution we first integrate by parts in $p$ in \Ref{3.2.32},
\eq{3.2.34}{  {\cal F}^{\rm L}_{\rm TM} &=-
   \int\frac{d\bm{k}}{(2\pi)^2}\int_0^\infty\frac{dp}{\pi}\,\frac{p}{\om}\,\frac{1}{e^{\om/T}-1}
   \, \delta^{\rm L}_{\rm TM}(p,\om).
}
We change the variable $k$ for $\om$,
\eq{3.2.35}{  {\cal F}^{\rm L}_{\rm TM} &=-\frac{1}{2\pi^2}\int_0^\infty  d\om\,
        \frac{1}{e^{\om/T}-1}\, h(\om),
}
where we defined
\eq{3.2.36}{h(\om) &=\int_0^\om dp\,p\,\delta^{\rm L}_{\rm TM}(p,\om).
}
In this case the integration over $p$ cannot be carried out analytically and one is left with the asymptotics of this function and numerical integration. For instance we note
\eq{3.2.37}{h(\om) &\raisebox{-4pt}{$=\atop \om\to0$} ~
\frac{4\om^3}{{\om_p}(e^{2{\om_p} L}-1)}+\dots\,.
}
Together with
\eq{3.2.38}{\delta^{\rm L}_{\rm TE}(p) &\raisebox{-4pt}{$=\atop p\to0$} ~ \frac{4p}{{\om_p}(e^{2{\om_p} L}-1)}+\dots\,.
}
which follows directly from \Ref{3.2.30} we get for $T\to0$
\eq{3.2.39}{
\Delta_T{\cal F}^{\rm L}_{\rm TE} &=\frac{-2\pi^2}{45 {\om_p}(e^{2{\om_p} L}-1)}T^4+\dots\,,\ \ \
\Delta_T{\cal F}^{\rm L}_{\rm TM} &=\frac{-2\pi^2}{15 {\om_p}(e^{2{\om_p} L}-1)}T^4+\dots\,.
}
For $T\to\infty$ we simply expand the logarithm in \Ref{3.2.33} resp.  the exponential in \Ref{3.2.35}. We get from  \Ref{3.2.33}
\eq{3.2.40}{
\Delta_T{\cal F}^{\rm L}_{\rm TE} &=\frac{T}{2\pi^2}\int_0^\infty d\om\,\om\,\ln\left(\frac{\om}{T}\right)
\delta^{\rm L}_{\rm TE}(\om)+\dots=d{\om_p}^2T+\dots\,,
}
where $d=\frac{1}{2\pi^2}\int_0^\infty d\om\,\om\,\ln\left({\om}\right)
\delta^{\rm L}_{\rm TE}(\om)_{|_{{\om_p}=1}}\simeq -0.0005936$. The integration \\
$\int_0^\infty d\om\,\om\,\delta^{\rm L}_{\rm TE}(\om)_{|_{{\om_p}=1}}\simeq 0$, which would result in a logarithmic contribution in \Ref{3.2.40}, gives zero within the precision of numerical integration.

From  \Ref{3.2.35} we get
\eq{3.2.41}{  {\cal F}^{\rm L}_{\rm TM} &=-\frac{T}{2\pi^2}\int_0^\infty  d\om\,
        \frac{1}{\om}\, h(\om)+\dots
            =c{\om_p}^2 T+\dots\,.
}
We get the same constant $d$ as from the TE polarization, again within the numerical precision.

In the Lifshitz part we do not need to do any subtraction since the growth with temperature of the free energy  does not exceed the first power. Therefore the entropy tends for $T\to\infty$ to a constant, $-d{\om_p}^2>0$, and is subleading as compared to the thickness independent part.
It should be mentioned, that the common way to get the $T\to\infty$-behavior of the free energy is to take the $l=0$-contribution in the Matsubara representation. As mentioned in \cite{klim15-92-042109}, this does not give a correct result.

The entropy of the Lifshitz part follows with \Ref{1.11} from \Ref{3.2.23} and \Ref{3.2.35} and reads
\eq{3.2.42}{ S^{\rm L}_{\rm TE} &=  \frac{1}{2\pi^2}\int_0^\infty d\om\,\om\,g\left(\frac{\om}{T}\right)
    \delta^{\rm L}_{\rm TE}(\om),
\nn\\   S^{\rm L}_{\rm TM} &=\frac{T^{-2}}{2\pi^2}\int_0^\infty  d\om\,\om
        \frac{e^{\om/T}}{\left(e^{\om/T}-1\right)^2}\, h(\om).
}
These are shown in Fig. 5.

\begin{figure}\label{fig3}
\includegraphics[width=7cm]{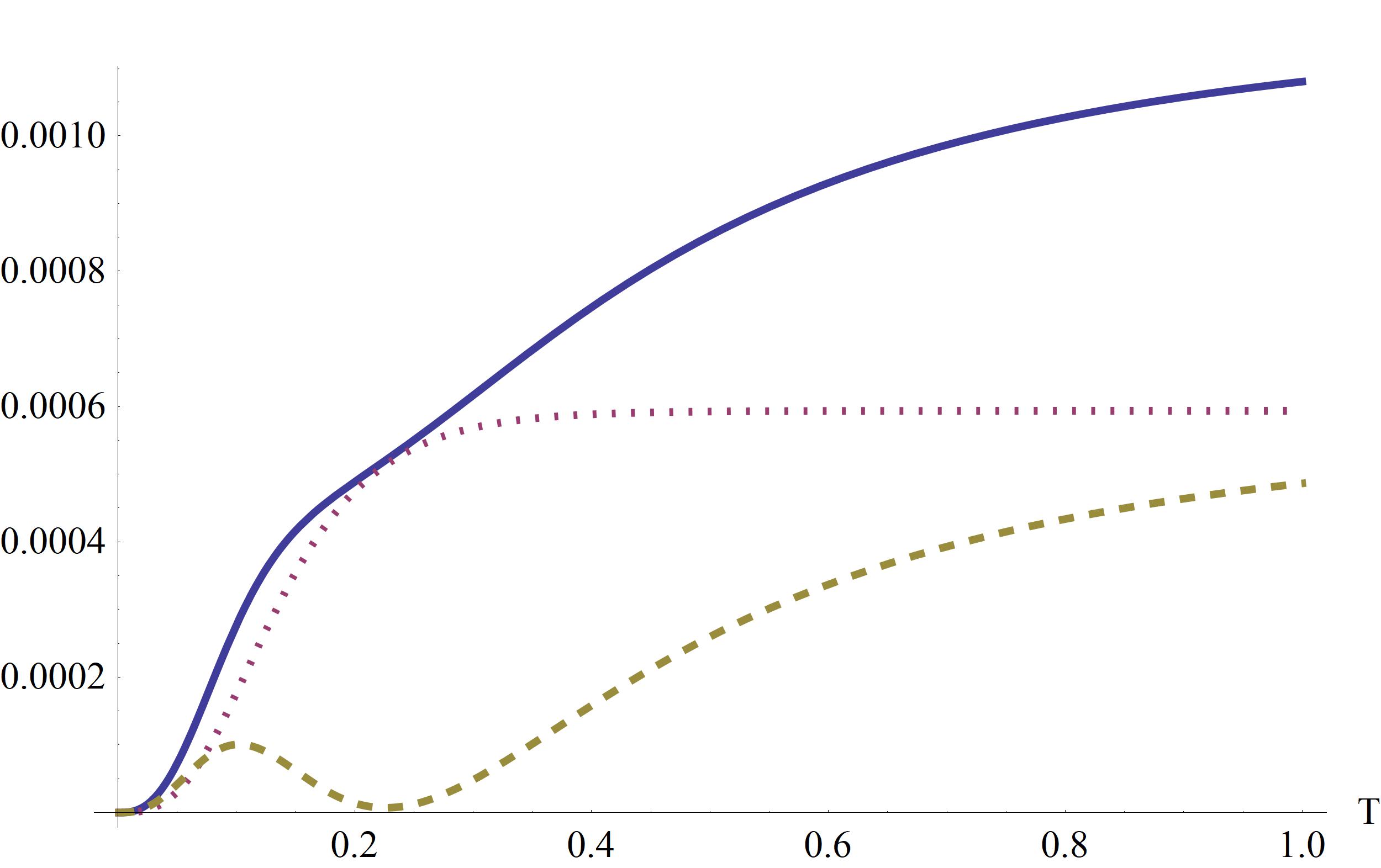} \ \ \ \ \  \includegraphics[width=7cm]{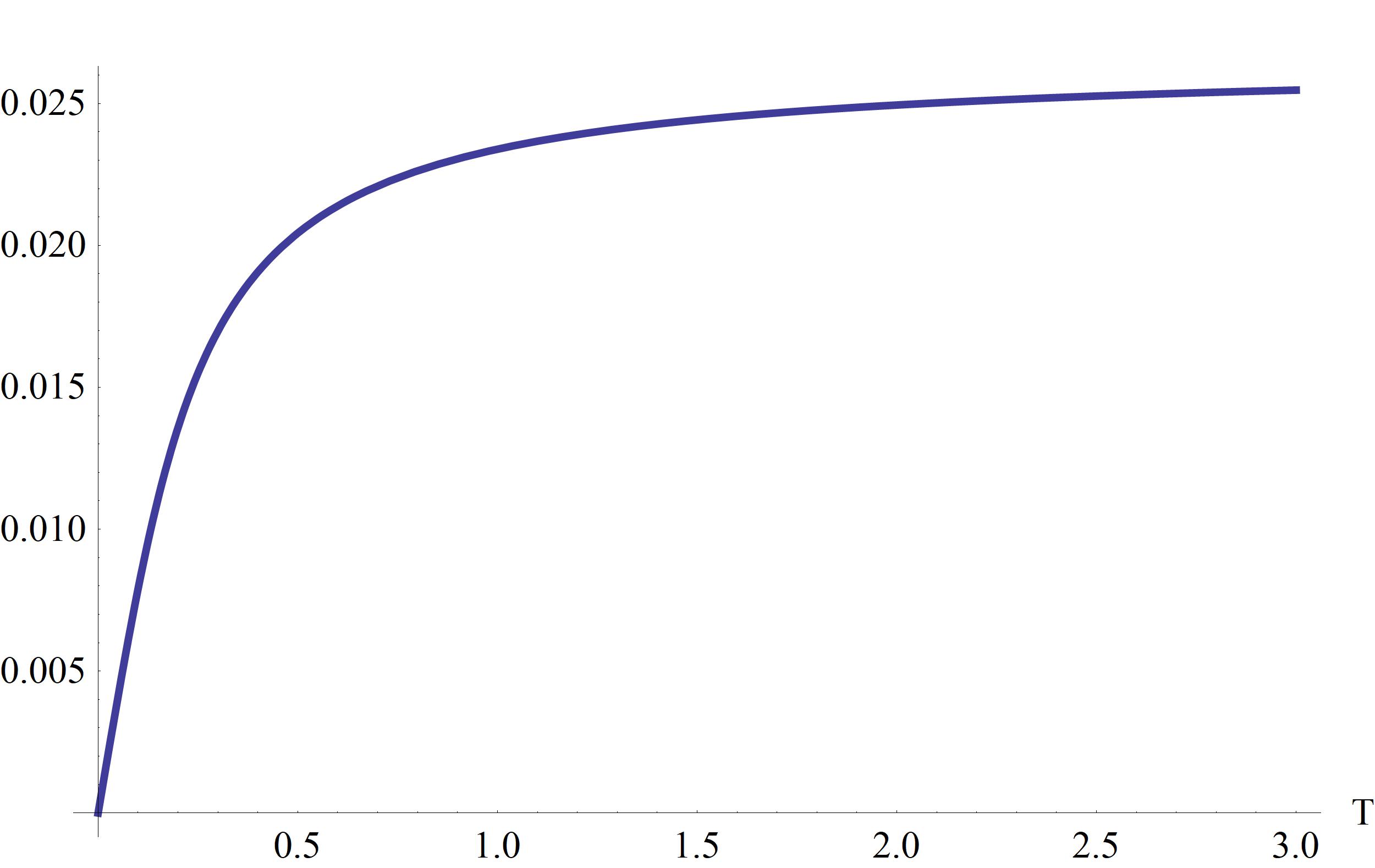}
\caption{In the left panel  the thickness dependent (Lifshitz) part of the entropy $S^{\rm L}$, \Ref{3.2.42}, of a  dielectric slab with ${\om_p}=1$ and $L=1$  as function of the temperature. The dotted line is the TE-contribution,
 the dashed   line is the TM-contribution and the solid line is the sum,
 ${S^{\rm L}}={S_{\rm TE}^{\rm L}}+{S_{\rm TM}^{\rm L}}$.
 In the right panel  the contribution ${S^{\rm exp}}^{\rm ~ subtr}$, \Ref{3.2.57}, to the entropy which is proportional to the thickness for $L=1$ and $\om_p=1$.}
\end{figure}

Now we come to the contribution from surface plasmons. These are poles of the transmission coefficient $t^{\rm L}_{\rm TM}$ \Ref{3.2.3} of the Lifshitz part. Their frequencies   $\om_{\rm sf}(k)$ are solutions of the equation \Ref{1.29}. These  enter the free energy through the first term in the square bracket in \Ref{1.10}.  For the dielectric slab this contribution is infinite since the frequency $\om_{\rm sf}(k)$ does not grow with $k$ as follows from the bound \Ref{1.31}. This is a property of the model. In applications it is not a problem since arbitrary high momenta $k$ are not supported by real materials and one has a Debye-cutoff.  However, for the complete free energy and the entropy, which we calculate, this would imply a dependence of the result on that cutoff. We do not go   into more details and keep this as an open question, speculating that like in the plasma sheet model the surface plasmons give a subleading contribution at high temperature.

%
\subsubsection{\label{T4.2.3}Contribution proportional to thickness ${\cal F}^{\rm exp} $}
We still have to consider the contributions from the exponential in the right side of \Ref{3.2.1}, i.e., $\Delta_T{\cal F}^{\rm exp} $ in eq. \Ref{3.2.5}. These are equal for the two polarizations. The corresponding factor in the transmission coefficient is simply $t=e^{i(q-p)L}$ with $q=\sqrt{p^2-\om_p^2}$, resulting in the phase  shift
\eq{3.2.50}{\delta(p)=\left\{\begin{array}{rl}
-pL,& (p<\om_p),  \\ (q-p)L,& (p>\om_p).    \end{array}\right.
}
The free energy takes the form
\eq{3.2.51}{\Delta_T{\cal F}^{\rm exp} &=
     \int\frac{d\bm{k}}{(2\pi)^2}   \left[ \int_0^{\om_p}\frac{dp}{\pi}\,
    T\ln\left(1-e^{-\om/T}\right)\frac{\pa}{\pa p}\left(-p\right)L
\right. \nn \\ &\left. ~~~~~~~ +\int^\infty_{\om_p}\frac{dp}{\pi}\,
    T\ln\left(1-e^{-\om/T}\right)\frac{\pa}{\pa p}\left(q-p\right)L \right] .
}
Carrying out the derivatives and rearranging the integrals we get
\eq{3.2.52}{\Delta_T{\cal F}^{\rm exp} &=
    L T\int\frac{d\bm{k}}{(2\pi)^2}   \left[ \int_0^{\infty}\frac{dp}{\pi}\,
    \ln\left(1-e^{-\om/T}\right)
+\int^\infty_{\om_p}\frac{dp}{\pi}\,\frac{p}{q}
     \ln\left(1-e^{-\om/T}\right) \right].
}
The first term in the square bracket results in an explicit integration. In the contribution from the second term we interchange the order of integrations and after that the $p$-integration can be carried out,
\eq{3.2.53}{\Delta_T{\cal F}^{\rm exp} &=\frac{\pi^2}{90}LT^4+
    \frac{L T}{2\pi^2}\int_{\om_p}^{\infty}d\om\,\om \sqrt{\om^2-\Om^2}
    \ln\left(1-e^{-\om/T}\right).
}
The behavior for $T\to0$ is given by the first term; the second is exponentially small.
The behavior for $T\to\infty$ can be obtained by a Mellin transform and reads
\eq{3.2.54}{\Delta_T{\cal F}^{\rm exp} \raisebox{-4pt}{$=\atop T\to\infty$}
    \frac{\om_p^2}{24}LT^2-\frac{\om_p^3}{12\pi}LT+\dots.
}
The $T^2$-contribution must be subtracted and we get
\eq{3.2.55}{{\Delta_T{\cal F}^{\rm exp}}^{\rm ~subtr} &=
\frac{\pi^2}{90}LT^4
- \frac{\om_p^2}{24}LT^2
+ \frac{L T}{2\pi^2}\int_{\om_p}^{\infty}d\om\,\om \sqrt{\om^2-\Om^2}
    \ln\left(1-e^{-\om/T}\right).
}
The entropy can be obtained using \Ref{1.11a} and reads
\eq{3.2.56}{{S^{\rm exp}}^{\rm~ subtr} &=
-\frac{2\pi^2}{45}LT^3
+ \frac{\om_p^2}{12}LT
+ \frac{L }{2\pi^2}\int_{\om_p}^{\infty}d\om\,\om \sqrt{\om^2-\Om^2}
   \, g\left(\frac{\om}{T}\right)
}
with the function $g$ defined in \Ref{1.11b}. Its asymptotic for $T\to\infty$ is
\eq{3.2.57}{{S^{\rm exp}}^{\rm ~ subtr} & \raisebox{-4pt}{$=\atop T\to\infty$}\frac{\om_p^3}{12\pi}L+\dots\,
}
and has no logarithmic contribution (which occurs in the next order of the expansion).

\section{\label{T5}Conclusions}
We calculated the entropy  for a flat plasma sheet and  for a dielectric slab. We had to subtract the contributions growing for high  temperature faster than the first power of $T$ (and $T\ln(T)$) as being unphysical. We use a representation of the free energy and the entropy in terms of real frequencies. This method of calculation allows to get an unambiguous result without any regularization. A special role play the surface plasmons which are present in the TM polarization. For the plasma sheet these give a well defined contribution to the entropy. For the dielectric slab the surface plasmons make the free energy and the entropy ill defined. We concluded that this behavior is a property of the model and restricted ourself to the calculation of the remaining contributions.

In Fig. 2, we observe negative entropy for the plasma sheet with intrinsic frequency $\om_0=0.85\Om_0/\sqrt{2}$, i.e., for a sheet of polarizable dipoles. In the case of $\om_0=0$, i.e., for a sheet filled with a charged fluid, which was considered in \cite{para17-96-085010}, the entropy is positive. We mention that the behavior of the entropy for $T\to\infty$ is logarithmic,
\eq{5.1}{ S&  \raisebox{-4pt}{$=\atop T\to\infty$}c(\om_0,\Om_0) \ln(T)+\dots\,,
}
where $c(\om_0,\Om_0)$ is some function which changes sign in dependence on the frequencies entering. The region where it takes negative values is narrow, $\frac{\Om_0}{\sqrt{2}}<\om_0   \lesssim         1.2\frac{\Om_0}{\sqrt{2}}$.

For the dielectric slab we observe a similar picture for $T\to\infty$ where the corresponding function is explicit, $c(\om_p)=-\frac{\om_p^2}{8\pi}$, \Ref{3.2.27}. This contribution results from the thickness independent part, whereas the Lifshitz part, which shows an interesting behavior (see Fig. 3 (left panel)) but stays positive in agreement with \cite{klim17-95-012130}.

We mention that negative entropy appears in the above examples for large temperature, whereas in the case of the spherical plasma shell \cite{bord1805.11241} it was observed for rather small temperature. So we conclude that this is a rather typical phenomenon for single bodies. Any further interpretation of these results we leave for future work.

\section*{Acknowledgements}
The author thanks G.L. Klimchitskaya and V.M. Mostepanenko for stimulating and helpful discussions. The author is also indebted to the organizers and participants of the {\it Casimir Effect Workshop 2018} in Trondheim  for the friendly atmosphere and stimulating discussions .


\begin{thebibliography}{10}

\bibitem{klim17-95-012130}
G.~L. Klimchitskaya and V.~M. Mostepanenko.
\newblock {Low-temperature behavior of the Casimir free energy and entropy of
  metallic films}.
\newblock {\em Phys. Rev. A}, 95:012130, 2017.

\bibitem{klim17-29-275701}
G.L. Klimchitskaya and V.M. Mostepanenko.
\newblock {Casimir free energy of dielectric films: classical limit,
  low-temperature behavior and control}.
\newblock {\em Journal of Physics: Condensed Matter}, 29:275701, 2017.

\bibitem{para17-96-085010}
Prachi Parashar, Kimball~A. Milton, K.~V. Shajesh, and Iver Brevik.
\newblock {Electromagnetic $\delta$-function sphere}.
\newblock {\em Phys. Rev.}, D96:085010, 2017.

\bibitem{milt17-96-085007}
Kimball~A. Milton, Pushpa Kalauni, Prachi Parashar, and Yang Li.
\newblock Casimir self-entropy of a spherical electromagnetic
  $\ensuremath{\delta}$-function shell.
\newblock {\em Phys. Rev. D}, 96:085007, Oct 2017.

\bibitem{geye05-72-022111}
B.~Geyer, G.~L. Klimchitskaya, and V.~M. Mostepanenko.
\newblock {Thermal corrections in the Casimir interaction between a metal and
  dielectric}.
\newblock {\em Phys. Rev. A}, 72:022111, 2005.

\bibitem{brev06-8-236}
Iver Brevik, Simen~A. Ellingsen, and Kimball~A. Milton.
\newblock {Thermal corrections to the Casimir effect}.
\newblock {\em New~J.~Phys.}, {8}:{236}, {2006}.

\bibitem{khus12-45-265301}
Nail~R Khusnutdinov.
\newblock {The thermal Casimir-–Polder interaction of an atom with a spherical
  plasma shell}.
\newblock {\em Journal of Physics A: Mathematical and Theoretical}, 45:265301,
  2012.

\bibitem{bord1805.11241}
M.~Bordag and K.~Kirsten.
\newblock On the entropy of a spherical plasma shell.
\newblock 2018.
\newblock {Arxiv 1805.11241}.

\bibitem{BV}
G.~Barton.
\newblock {Casimir effects for a flat plasma sheet: I. Energies}.
\newblock {\em J.~Phys.~A: Math.~Gen.}, 38:2997--3019, 2005.

\bibitem{bord05-38-11027}
M.~Bordag, I.~G. Pirozhenko, and V.~V. Nesterenko.
\newblock Spectral analysis of a flat plasma sheet model.
\newblock {\em J. Phys.}, A38:11027, 2005.

\bibitem{fett73-81-367}
A.~L. Fetter.
\newblock {Electrodynamics of a Layered Electron-Gas.1. Single Layer}.
\newblock {\em Ann.~Phys.}, 81:367--393, 1973.

\bibitem{BKMM}
M.~Bordag, G.~L. Klimchitskaya, U.~Mohideen, and V.~M. Mostepanenko.
\newblock {\em Advances in the Casimir Effect}.
\newblock Oxford University Press, Oxford, 2009.

\bibitem{bord95-28-755}
M.~Bordag.
\newblock {Vacuum Energy in Smooth Background Fields}.
\newblock {\em J. Phys.}, A28:755--766, 1995.

\bibitem{dowk78-11-895}
J~S Dowker and G~Kennedy.
\newblock Finite temperature and boundary effects in static space-times.
\newblock {\em J.~Phys.~A: Math.~Gen.}, 11:895, 1978.

\bibitem{vass03-388-279}
D.V. Vassilevich.
\newblock Heat kernel expansion: User's manual.
\newblock {\em Phys. Rept.}, 388:279--360, 2003.

\bibitem{kirs01b}
K.~Kirsten.
\newblock {\em Spectral Functions in Mathematics and Physics}.
\newblock Chapman\&Hall/CRC, Boca Raton, FL, 2001.

\bibitem{bart13-15-063028}
G.~Barton.
\newblock Casimir effects in monatomically thin insulators polarizable
  perpendicularly: nonretarded approximation.
\newblock {\em New~J.~Phys.}, 15:063028, 2013.

\bibitem{klim15-91-045412}
G.~L. Klimchitskaya and V.~M. Mostepanenko.
\newblock {Comparison of hydrodynamic model of graphene with recent experiment
  on measuring the Casimir interaction}.
\newblock {\em Phys. Rev.}, B91:045412, 2015.

\bibitem{intr05-94-110404}
F.~Intravaia and A.~Lambrecht.
\newblock {Surface plasmon modes and the Casimir energy}.
\newblock {\em Phys.~Rev.~Lett.}, 94:110404, 2005.

\bibitem{bord05-39-6173}
M.~Bordag.
\newblock {The Casimir effect for thin plasma sheets and the role of the
  surface plasmons}.
\newblock {\em J.~Phys.~A: Math.~Gen.}, 39:6173--6185, 2006.

\bibitem{bord12-85-025005}
M.~Bordag.
\newblock {Electromagnetic Vacuum Energy for two Parallel Slabs in Terms of
  Surface, Wave Guide and Photonic Modes}.
\newblock {\em Phys.~Rev.~D}, {85}:{025005}, 2012.

\bibitem{klim15-92-042109}
G.~L. Klimchitskaya and V.~M. Mostepanenko.
\newblock {Casimir free energy of metallic films: Discriminating between Drude
  and plasma model approaches}.
\newblock {\em Phys. Rev. A}, 92:042109, 2015.

\end{thebibliography}
\end{document}